\documentclass[conference]{IEEEtran}
\IEEEoverridecommandlockouts
\usepackage{amsmath}
\usepackage{acronym}
\usepackage{multirow}
\usepackage{tabularray}
\usepackage{cite}
\usepackage{needspace}
\usepackage{pifont}
\usepackage[acronym]{glossaries}
\acrodef{LLM}[LLM]{Large Language Model}
\acrodef{NBPU}[NBPU]{Near-Bank Processing Unit}
\acrodef{PIM}[PIM]{Processing In-Memory}
\acrodef{CXL}[CXL]{Compute Express Link}
\acrodef{ISL}[ISL]{Input Sequence Length}
\acrodef{SSM}[SSM]{State Space Model}
\acrodef{OSL}[OSL]{Output Sequence Length}
\usepackage{footnote}
\usepackage{footnote}
\usepackage{algorithm}
\usepackage{algpseudocode}

\makeatletter
\let\CL@fs@ruled\fs@ruled
\renewcommand{\fs@ruled}{%
  \CL@fs@ruled
  \let\CL@fs@pre\@fs@pre
  \def\@fs@pre{\vskip 5pt\relax\CL@fs@pre}%
}
\makeatother

\usepackage{booktabs}
\usepackage{graphicx}
\usepackage{newtxtext,newtxmath}
\usepackage{tabularx}
\usepackage[flushleft]{threeparttable}
\usepackage{libertine}
\usepackage{circledsteps}
\pgfkeys{/csteps/inner color=white}
\pgfkeys{/csteps/fill color=black}
\usepackage{enumitem}
\usepackage{tikz}
\usetikzlibrary{patterns,positioning,calc,fit}
\usepackage{xcolor}
\usepackage{enumitem}
\usepackage{orcidlink}
\hypersetup{hidelinks}

\begin{document}
\title{On Design Principles for Efficient Heterogeneous DRAM-PIM-GPU Systems}
\author{\IEEEauthorblockN{Corey Lammie\orcidlink{0000-0001-5564-1356}\IEEEauthorrefmark{1}\IEEEauthorrefmark{2}\thanks{\hspace{-1em}\rule{3cm}{0.5pt} \newline \textcopyright  \hspace{1pt} 2026 IEEE. Personal use of this material is permitted. Permission from IEEE must be obtained for all other uses, in any current or future media, including reprinting/republishing this material for advertising or promotional purposes, creating new collective works, for resale or redistribution to servers or lists, or reuse of any copyrighted component of this work in other works.}, Hadjer Benmeziane\orcidlink{0000-0002-5259-0749}\IEEEauthorrefmark{1}, William Andrew Simon\orcidlink{0000-0001-7357-7204}\IEEEauthorrefmark{1}, Irem Boybat\orcidlink{0000-0002-4255-8622}\IEEEauthorrefmark{1}}
	\IEEEauthorblockA{
		\IEEEauthorrefmark{1}IBM Research, Säumerstrasse 4, 8803 R\"{u}schlikon, Switzerland, \IEEEauthorrefmark{2}Email: corey.lammie@ibm.com}
}

\maketitle

\begin{abstract}
	Heterogeneous DRAM-based processing-in-memory (PIM)-GPU systems promise significant efficiency gains for decode-phase large language model (LLM) inference, particularly in long-output generation, yet current design practices overlook critical factors that determine real-world performance. Through systematic evaluation of diverse architectures and workloads (OPT-7B/70B, Mamba2-2.7B/70B), we reveal three fundamental design principles: (i) static power consumption (DRAM leakage, refresh, and GPU idle power) can dominate the efficiency calculus, causing dynamic-only models to overestimate tokens/s/W by up to 3.85$\times$ for realistic deployments (Mamba2-2.7B, batch size 1, 128 input tokens, and 2,048 output tokens); (ii) decoding performance is monotonically non-decreasing with channel count across all evaluated models and workloads, generally plateauing at high channel counts for low-batch workloads; under a fixed-capacity sweep, all models instead share a common near-optimal hierarchy configuration, with substantially larger misconfiguration penalties for attention-based models; (iii) workload mapping strategies provide bounded improvements (up to 14.0\%/17.4\% kernel-level latency/energy reduction, up to 5.6\% end-to-end gain) and are not primary bottlenecks. Significant efficiency gains require system-wide co-optimization. These principles provide design-space guidance for architects designing the next generation of memory-accelerated LLM systems.
\end{abstract}


\glsresetall
\section{Introduction}\label{sec:introduction}
\begin{figure}[!t]
	\centerline{\includegraphics[width=1\columnwidth]{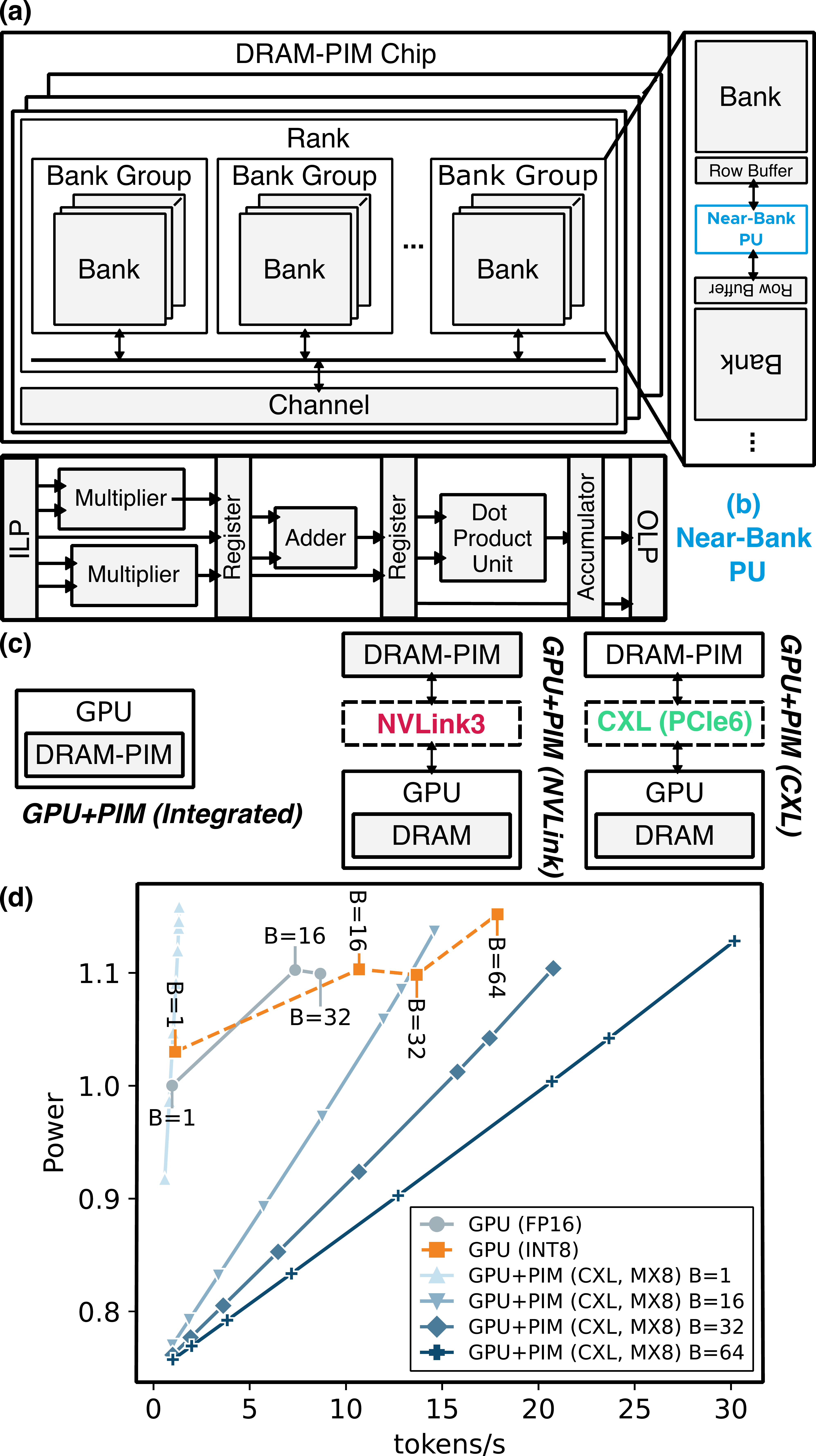}}
	\caption{(a) DRAM-PIM chip architecture with reconfigurable channels, ranks, bank groups, and banks. (b) Near-Bank Processing Unit (NBPU) connected to pairs of banks. (c) Heterogeneous DRAM-PIM-GPU systems considered. (d) Decoding performance vs.\ channel count for GPU+PIM (CXL) (OPT-7B, ISL=2048, OSL=2048); GPU baseline ($\times$1 A100 (80\,GiB)) varies batch size.}
	\label{fig:key_conceptual}
\end{figure}

\acp{LLM} push modern hardware against memory-bandwidth bottlenecks: GPUs excel at compute, but attention, recurrent state updates, and other memory-intensive operations stall on bandwidth.
\ac{PIM}, particularly DRAM-based PIM (DRAM-PIM), offers a promising path forward by bringing computation closer to data and reducing costly data movement~\cite{khan2024landscapecomputenearmemorycomputeinmemoryresearch}.
Due to the high maturity of DRAM and low adoption barrier of DRAM-PIM, heterogeneous systems that co-locate or integrate DRAM-PIM units with GPUs, such as those depicted in Fig.~\ref{fig:key_conceptual}a-c, are emerging as an attractive platform for decode-phase LLM acceleration in long-output generation~\cite{Gu2025}.

Although recent work highlights the potential of using DRAM-PIM for \ac{LLM} execution, the broader architectural implications remain insufficiently understood. Key design factors, namely DRAM hierarchy organization, static power, workload mapping, and interconnect choice, interact in non-obvious ways. For instance, adding DRAM channels can yield \textit{diminishing returns} when workloads lack sufficient memory-level parallelism (Fig.~\ref{fig:key_conceptual}d), and static power from idle components can negate PIM throughput gains entirely at low batch sizes.
Through comprehensive system-level evaluation, we derive three actionable design principles that provide concrete guidance for designing efficient heterogeneous memory systems:
\begin{itemize}[noitemsep,topsep=2pt,leftmargin=10pt]
	\item We quantify the impact of static power (DRAM leakage, refresh, GPU idle) on system-level efficiency, showing that neglecting it overestimates tokens/s/W by up to 3.85$\times$ and fundamentally shifts optimal design points;
	\item We demonstrate that decoding performance is monotonically non-decreasing with channel count across all evaluated models and workloads, generally plateauing at high channel counts for low-batch workloads. Under a fixed-capacity sweep, all models instead share a common near-optimal region; the risk is an approximately 2$\times$ larger fractional misconfiguration penalty for attention-based than SSM-based models;
	\item Finally, we present RowLocalChunk, a locality-aware workload mapping, showing that it reduces kernel-level latency/energy by up to 14.0\%/17.4\%, while end-to-end (mapping-level) gains remain bounded (up to 5.6\%).
\end{itemize}

\section{Background and Motivation}\label{sec:background}
DRAM-based \ac{PIM} extends commodity memory with low-cost compute capability~\cite{khan2024landscapecomputenearmemorycomputeinmemoryresearch}, with feasibility demonstrated in manufactured silicon (Samsung, SK Hynix, UPMEM). In heterogeneous systems, GPUs can handle compute-bound kernels while \ac{PIM} units accelerate bandwidth-limited operations such as KV-cache accesses and state updates~\cite{Kim2025, Park2024}, aligning naturally with modern \acp{LLM}.
Achieving system-level efficiency, however, depends jointly on static power (e.g., DRAM leakage/refresh, GPU idle power during PIM execution), DRAM hierarchy organization (e.g., channels, ranks, bank groups, banks), workload partitioning/mapping (fixed vs.\ dynamic~\cite{He2025}), data layout~\cite{Wang2024}, and GPU--PIM interconnect choice (e.g., integrated vs.\ \ac{CXL}/NVLink-attached~\cite{Gu2025}), each shaping latency, bandwidth, and power differently. Existing evaluation tools address only subsets of these factors: NicePIM~\cite{Wang2024} and Pimba~\cite{Kim2025} target workload mapping; PAPI~\cite{He2025} and AttAcc~\cite{Park2024} additionally model interconnects; none jointly capture static power alongside hierarchy exploration for LLM decode. We evaluate all of these factors jointly across OPT and Mamba2 decode workloads.

\section{Experimental Setup}
\begin{figure}[!t]
	\centerline{\includegraphics[width=1\columnwidth]{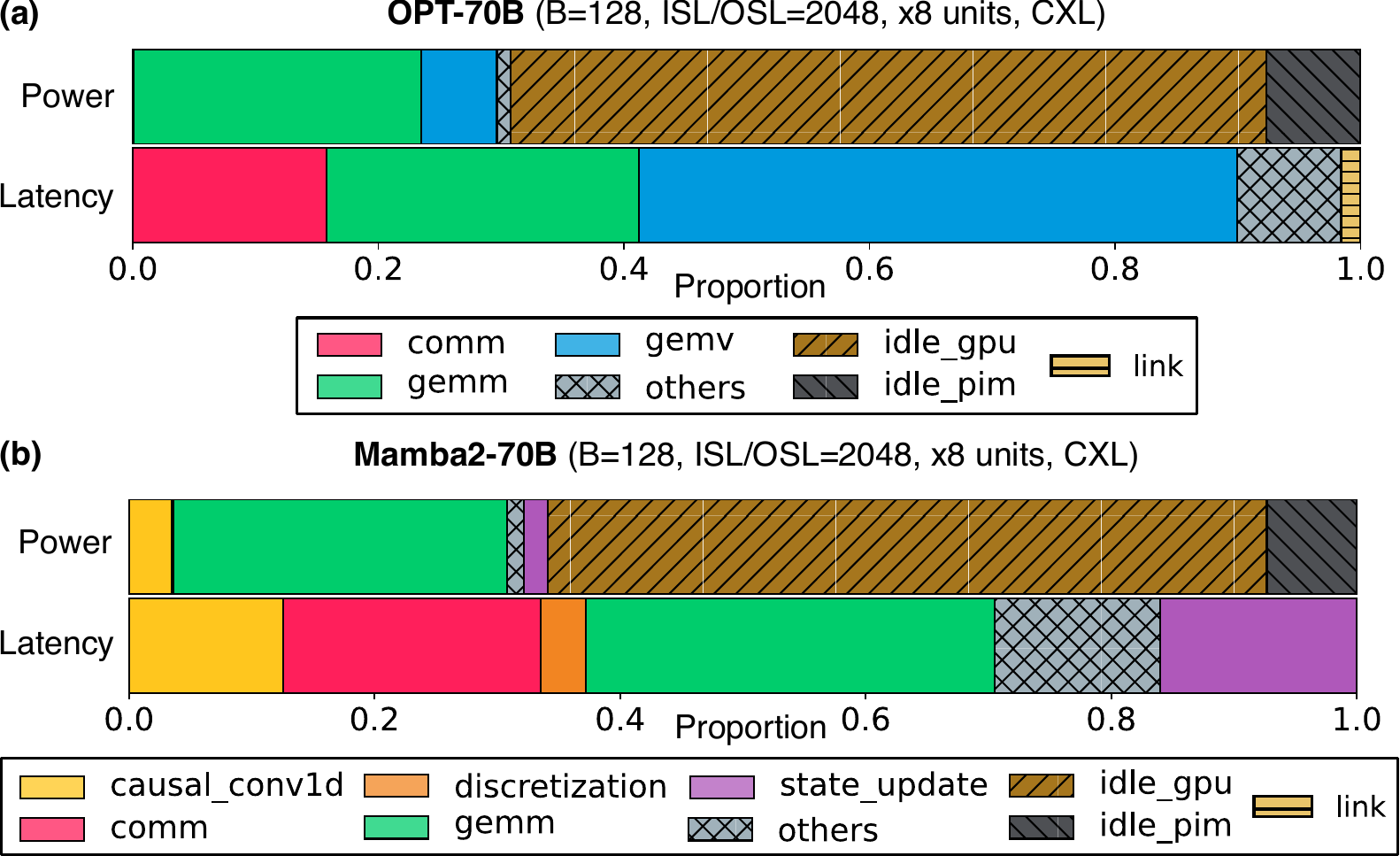}}
	\caption{Power and latency breakdown for OPT-70B and Mamba2-70B ($\times$8 units, B=128, ISL/OSL=2048, GPU+PIM (CXL)).}
	\label{fig:breakdown}
\end{figure}

We study heterogeneous GPU--DRAM-PIM systems across hierarchy organizations, workload mappings, and interconnect options.
We instantiate a representative \ac{NBPU} design following Pimba~\cite{Kim2025}/AttAcc~\cite{Park2024}: 640 NBPUs operating at 378\,MHz per unit, one per two banks (for the default 40-channel/1{,}280-bank hierarchy), each supporting MX8 dot-product/multiply--accumulate and accumulation per the cited architectures --- an architectural capability we adopt from these proposals rather than independently re-verify at the ALU level, since this simulator tracks NBPU activity only as an opaque per-cycle compute-event count.
Active NBPU energy is accounted per executed compute event, scaling with operation count, while \texttt{idle\_pim} (Fig.~\ref{fig:breakdown}) is the co-located PIM-HBM's precharged-standby background power plus its average refresh power (capacity-normalized as above), charged only to CXL/NVLink co-located configurations and never to integrated DRAM-PIM, which shares the GPU's own already-installed HBM.

Our goal is comparative design-space evaluation rather than cycle-exact reproduction of a commercial GPU. We therefore calibrate static GPU power using NVML idle measurements, DRAM leakage and refresh using prior experimental DRAM characterization~\cite{Ghose2018} and DRAMPower-style parameters~\cite{Steiner2025}, and interconnect latency/bandwidth using published NVLink/CXL values.
We adapt and extend ramulator2~\cite{Luo2023} (memory) plus attacc\_simulator~\cite{Park2024} and Pimba~\cite{Kim2025} (system) to model static power (DRAM refresh/leakage), flexible operation mapping, and both integrated and co-located DRAM-PIM--GPU configurations across interconnect types. This captures GPU idle power during PIM execution, transfer/compute contention, and hierarchy-mapping interplay.

Each system pairs one or more NVIDIA A100 GPUs (80\,GiB HBM2e, peak memory bandwidth 1{,}935\,GB/s, 5 HBM stacks, 108 SMs~\cite{Choquette2021}) with DRAM-PIM.
We consider three integration strategies, as introduced in Section~\ref{sec:background}: integrated DRAM-PIM sharing the GPU memory bus, and co-located DRAM-PIM connected via CXL (PCIe 6.0) or NVLink3.
We hold the physical-channel transfer rate at 3,024\,MT/s; aggregate DRAM-PIM bandwidth scales with the configured channel count.
HBM2-vs-HBM2e differences fall within our $\pm$20\% sensitivity envelope (\S\ref{sec:static}).
The DRAM is configured with a refresh interval of 3.9\,$\mu$s, matching Pimba's reported HBM $t_{\mathrm{REFI}}$~\cite{Kim2025}, and a refresh energy of 4.89\,nJ per bank per event (0.72--4.89\,nJ sensitivity) -- our own density/capacity-scaled modeling estimate, derived using the DRAMPower methodology~\cite{Steiner2025} rather than a value either work reports directly.
Where per-unit PIM capacity is fixed but distributed across different numbers of physical banks (Fig.~\ref{fig:tsne}), we scale refresh energy per bank event linearly with per-bank capacity:
\[
	E_{\mathrm{ref,bank}}
	=4.89\,\mathrm{nJ}\times
	\frac{C_{\mathrm{bank}}}{64\,\mathrm{MiB}},
\]
where 64\,MiB is the per-bank capacity of the reference 1,280-bank organization. At fixed $t_{\mathrm{REFI}}$, aggregate refresh power therefore tracks the fixed per-unit capacity rather than the number of banks. A separate, physically co-located PIM-HBM unit (CXL/NVLink-attached) additionally draws a flat 8.41\,W/unit precharged-standby background power (2.10--8.41\,W sensitivity), our own capacity-scaled estimate for an 80\,GiB HBM stack derived from the DDR3L background-power characterization in~\cite{Ghose2018} (no HBM-specific value is directly reported); this value is kept separate from the A100 (80\,GiB) system-level power measurement, and this background term is \emph{not} charged for integrated DRAM-PIM, which shares the GPU's own already-installed HBM. Context-active GPU power is modeled at a nominal 80\,W per A100 board, including installed HBM, based on our observed NVIDIA Management Library (NVML) telemetry. We additionally validate, per GPU, that context power plus dynamic energy over time never exceeds the A100 PCIe 300\,W board limit~\cite{Choquette2021}. Table~\ref{tab:power_params} summarizes all static-power and interconnect parameters (main/nominal values; sensitivity ranges are evaluated in Table~\ref{tab:sensitivity}, \S\ref{sec:static}).

\begin{table}[!t]
	\caption{Static-power and interconnect parameters. Refresh energy $\times$ 1{,}280 banks / refresh interval $\approx$ 1.60\,W/unit nominal refresh power.}
	\label{tab:power_params}
	\centering
	\scriptsize
	\begin{threeparttable}
		\begin{tabular}{lr}
			\toprule \toprule
			\textbf{Component} & \textbf{Value}                         \\
			\midrule
			A100 context-active (board, incl.\ HBM)
			                   & 80\,W                                  \\
			PIM-HBM background power (co-located only)
			                   & 8.41\,W/unit~\cite{Ghose2018}          \\
			Refresh energy
			                   & 4.89\,nJ/bank/event~\cite{Steiner2025} \\
			Refresh interval
			                   & 3.9\,$\mu$s~\cite{Kim2025}             \\
			NVLink3 BW (directional)
			                   & 300\,GB/s~\cite{Choquette2021}         \\
			NVLink3 latency
			                   & 1.0\,$\mu$s~\cite{Park2024}            \\
			CXL over PCIe 6.0 BW (directional)
			                   & 120\,GB/s                              \\
			\bottomrule \bottomrule
		\end{tabular}
	\end{threeparttable}
\end{table}

Following common practice~\cite{Kim2025}, prefill operations are executed entirely on the GPU(s), as they are predominantly compute-bound. Our analysis targets decode, which dominates latency in long-output chat/code-generation deployments. Self-attention and state update operations, which involve high-volume, fine-grained data access patterns naturally suited to in-memory computation, are offloaded to the DRAM-PIM chip(s). All remaining operations execute on the GPU(s). DRAM-PIM computation uses the MX8 format~\cite{DarvishRouhani2023}, storing KV-cache/state at 1.0\,byte/value (metadata included) for both performance and capacity accounting. We additionally report an idealized GPU-INT8 baseline (``GPU+Q'') alongside GPU-FP16: KV-cache/state uses 1.0625\,bytes/value for capacity accounting (1\,byte plus one BF16 scale factor per 32-value group) but the idealized 1.0\,byte/value rate for traffic/latency/energy, with no quantization overhead.
Capacity is checked independently per physical allocation domain (weights, activations, KV/state); co-located DRAM-PIM adds a separate 80\,GiB PIM-HBM pool per unit, while integrated DRAM-PIM shares the GPU's own HBM; an equal-unit, not iso-area/iso-cost/fixed-power-budget, comparison, for which integrated results provide the shared-capacity reference. The $\leq$6.25\% storage-density gap between GPU-INT8 and MX8 is a second-order effect next to this topology-level capacity difference, but can determine feasibility outright for workloads near the 80\,GiB boundary (e.g., OPT-7B at B=128, ISL=2048, OSL=128).
Transfer latencies ($k$ bytes) over NVLink3 and CXL/PCIe\,6.0 are modeled with a canonical fixed-latency-plus-bandwidth link model:
\[
	t_s(k) = L_s + k/B_s, \quad s \in \{\text{NVLink}, \text{CXL}\},
\]
\noindent with NVLink directional bandwidth $B_{\text{NVLink}}\!=\!300$\,GB/s and latency $L_{\text{NVLink}}\!=\!1.0\,\mu$s, and CXL directional bandwidth $B_{\text{CXL}}\!=\!120$\,GB/s and fixed latency $L_{\text{CXL}}\!=\!4l_p\!+\!2l_e\!+\!l_s\!=\!180$\,ns, with port latency $l_p\!=\!25$\,ns, PCIe latency $l_e\!=\!30$\,ns, switch latency $l_s\!=\!20$\,ns. This replaces an earlier AttAcc-derived empirical curve fit for NVLink calibrated to a bidirectional-aggregate 600\,GB/s figure that does not correspond to a one-way transfer rate. GPU$\leftrightarrow$GPU tensor-parallel collective communication (used by every 8-GPU/70B result in this paper) is modeled separately from the GPU$\leftrightarrow$PIM link above, via a canonical recursive halving/doubling all-reduce over the same directional NVLink bandwidth: $t_{\text{AR}} = 2\log_2(P)L_{\text{AR}} + 2\frac{P-1}{P}\cdot\frac{k}{B_{\text{NVLink}}}$, with per-round collective latency $L_{\text{AR}}\!=\!6.06\,\mu$s chosen to preserve the empirical intercept of the AttAcc curve fit it replaces, which is retained in our codebase only as a legacy comparison mode and does not produce any result reported in this paper. Our 8-GPU results therefore combine PCIe-A100 board specifications (1{,}935\,GB/s HBM2e, 300\,W board limit) with an NVLink3/NVSwitch-style all-to-all collective network.

\begin{algorithm}[!t]
	\caption{RowLocalChunk mapping.}
	\label{alg:genmap_compact}
	\begin{algorithmic}[1]
		\State $chunk_r \!\gets\! \lfloor \text{atom\_sz}/\text{elem\_w} \rfloor$;\; $chunk_c \!\gets\! \lfloor \text{page\_sz}/\text{atom\_sz} \rfloor$
		\State $N_r \!\gets\! \lceil d_{\text{model}}/chunk_r \rceil \!\cdot\! B$;\; $N_c \!\gets\! \lceil L_{\max}/chunk_c \rceil$
		\State $\text{quota} \gets \text{distribute}(N_r N_c,\, U)$ \Comment{largest-remainder}
		\State Init $\text{row\_own}[r], \text{col\_own}[c] \gets \varnothing$;\; $N_{core} \!\gets\! \max(1, \text{heads} \!\times\! B)$
		\ForAll{$(r,c)$ in row-major order}
		\State $C \gets \{u \in U \mid \text{quota}[u] > 0\}$
		\State $chosen \gets \begin{cases}
		\min(C \cap \text{row\_own}[r]), & \text{if } \neq \emptyset\\
		\min(C \cap \text{col\_own}[c]), & \text{else if } \neq \emptyset\\
		\arg\min_{u \in C} \text{load}[u], & \text{else}
		\end{cases}$
		\State Assign $(r,c) \!\to\! chosen$; update quota, ownership
		\State Core: reuse bank's core, else $idx\!\!\mod N_{core}$ (round-robin)
		\EndFor
		\State \Return per-unit chunk mapping $\mathcal{M}$
	\end{algorithmic}
\end{algorithm}

\begin{figure}[!t]
	\centering
	\centerline{\includegraphics[width=0.385\textwidth]{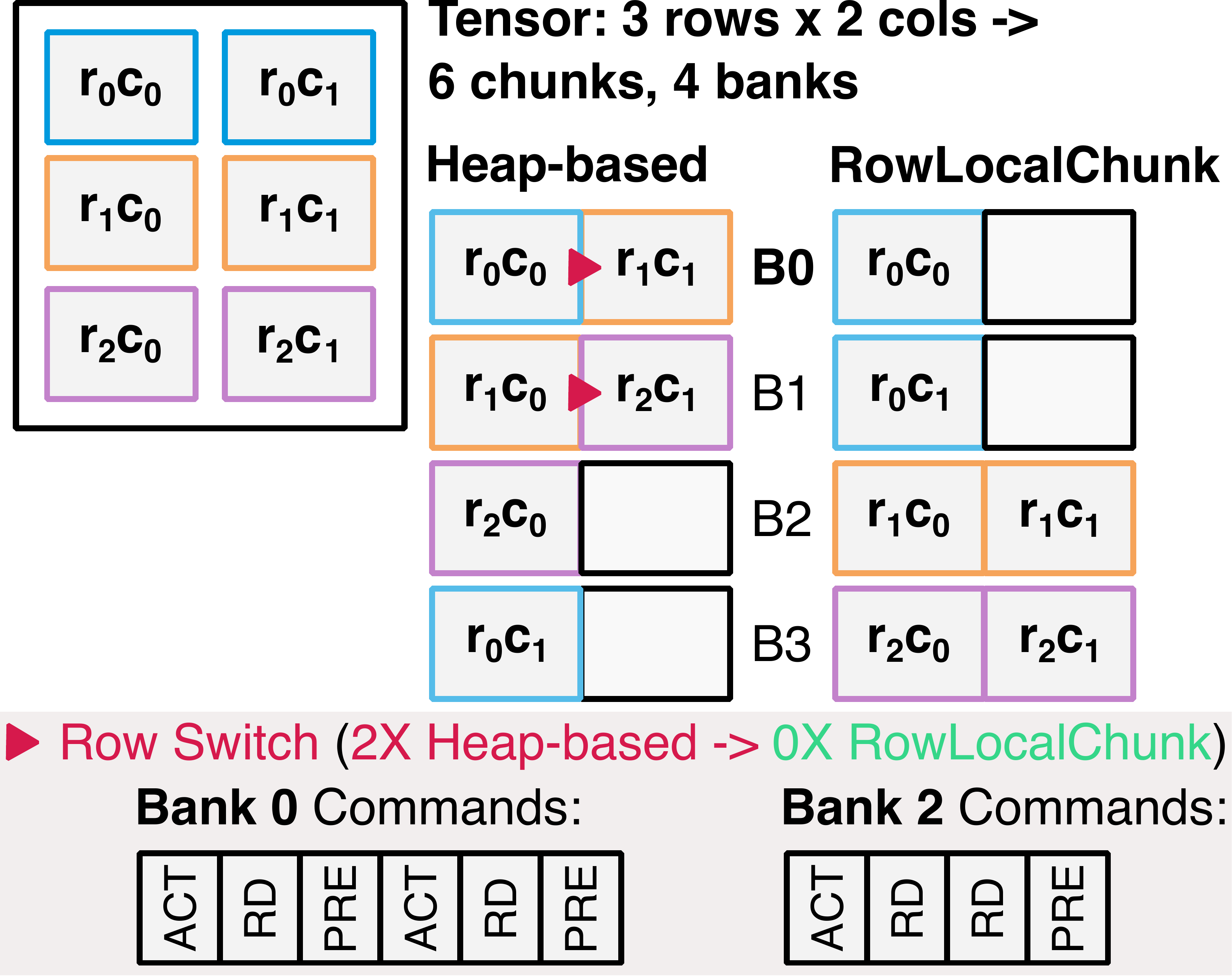}}
	\vspace{-0.5em}
	\caption{Allocation of 6 tensor chunks ($3$ rows $\times$ $2$ columns) to 4 banks. Heap-based (Pimba) assigns chunks to globally least-filled rows, potentially fragmenting tensor rows across banks. In contrast, RowLocalChunk performs a quota-based triage (row-owner $\to$ col-owner $\to$ least-loaded).}
	\label{fig:mapping_illustration}
\end{figure}

\begin{figure*}[!t]
	\centerline{\includegraphics[width=1\textwidth]{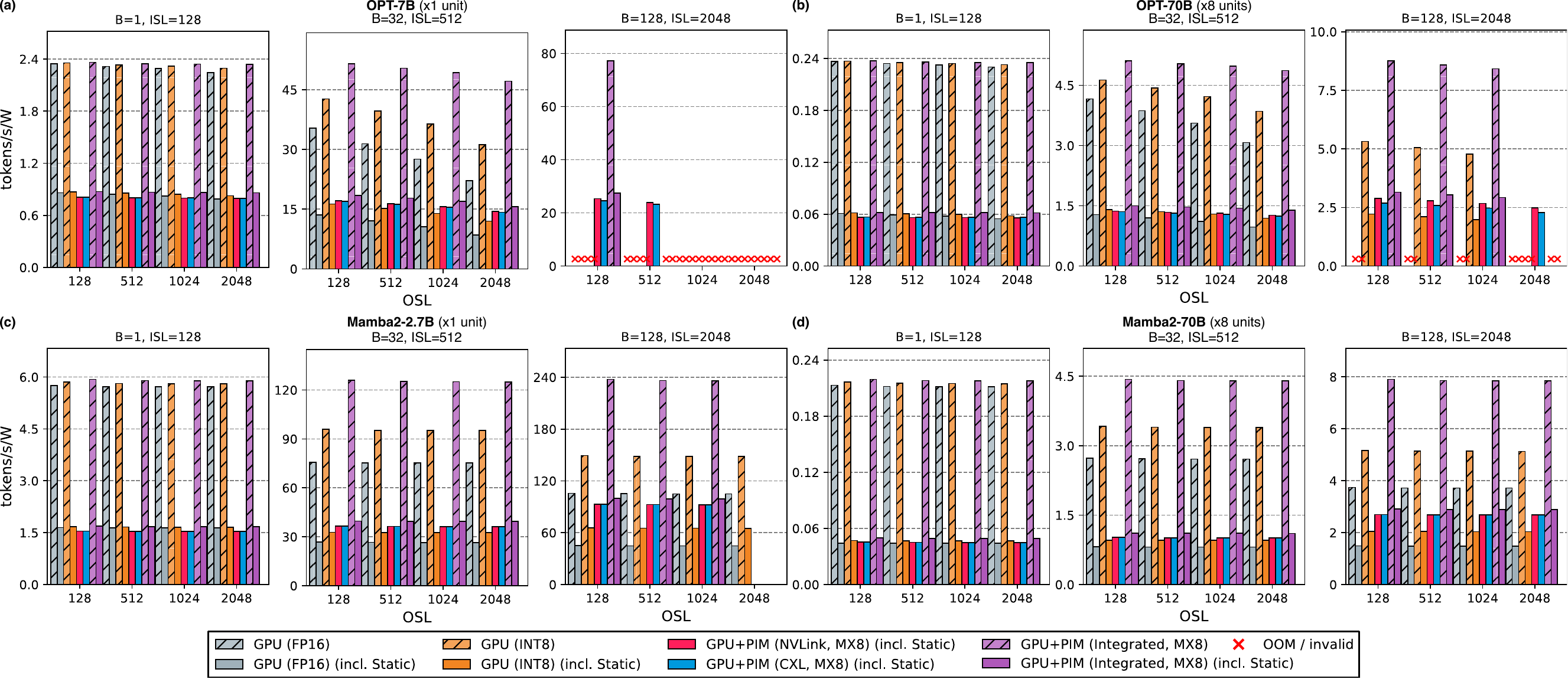}}
	\vspace{-0.5em}
	\caption{Decoding performance for (a) OPT-7B, (b) OPT-70B ($\times$8), (c) Mamba2-2.7B, (d) Mamba2-70B ($\times$8). DRAM hierarchy: $ch\!=\!40, pseudochannel\!=\!2, bg\!=\!4, bank\!=\!4$.}
	\label{fig:static}
\end{figure*}

We evaluate two standard decoder-based transformers with KV caching, representative of GPT-like \acp{LLM}, OPT-7B and OPT-70B~\cite{Zhang2022}, and two \acp{SSM}: Mamba2-2.7B using the original Mamba2 configuration~\cite{Dao2024}, and a scaled Mamba2-70B model with 192 decoder blocks, hidden dimension 7680, 80 heads, head dimension 384, state dimension 192, and 2-byte data elements.
For the smaller networks ($\leq7B$), a single GPU-DRAM-PIM unit is used. For the larger ones, 8 units are used. Each unit is a tensor-parallel shard pairing one A100 (80\,GiB) with a DRAM-PIM chip; weights, activations, and KV cache are sharded uniformly. At B=128 and ISL=OSL=2048, the OPT-style 70B FP16 KV cache exceeds the aggregate 640-GiB GPU capacity, while its one-byte representation alone reaches 640\,GiB before accounting for headroom; we therefore mark infeasible baselines as OOM. The GPU-only baseline uses the same number of A100 GPUs as the heterogeneous configuration; total memory capacity is not matched for co-located PIM, while integrated PIM provides the shared-capacity reference. Unless otherwise stated, a fixed DRAM hierarchy of 40 channels, 2 pseudochannels, 4 bank groups, and 4 banks is used to facilitate comparison with related work. For all baselines, the equivalent number of GPUs (to units) is used.

\subsection{Workload Mapping Strategy}\label{sec:mapping}
As baseline, we use Pimba's heap-based allocator~\cite{Kim2025}, which assigns chunks to the least-filled DRAM row via a min-heap. This approach is simple and load-balanced, but it scatters adjacent chunks across distant banks/rows, producing high row-switching and nondeterministic layouts.
We introduce \textit{RowLocalChunk} (Algorithm~\ref{alg:genmap_compact})\footnote{Notation: $\text{atom\_sz}\!=\!32$\,B is the storage size of one 32-element MX8 atom~\cite{DarvishRouhani2023}, $\text{elem\_w}\!=\!1$\,B/element, and $\text{page\_sz}\!=\!1024$\,B is the row-buffer size; $d_{\text{model}}/L_{\max}$ hidden dim/max length; $B$ batch; $U$ PIM units. Largest-remainder: $\lfloor N_r N_c/|U|\rfloor$ baseline plus residue-ranked top-ups}, a simple deterministic locality-aware mapping prioritizing spatial locality while preserving global balance. Per-unit integer quotas are computed via largest-remainder; chunks are traversed in row-major order and assigned via three-stage triage: same-row owner, same-column owner, then least-loaded bank. This mitigates the scattered allocations of heap-based methods (Fig.~\ref{fig:mapping_illustration}).

\section{Results}

\subsection{Static Power Consumption}\label{sec:static}

\begin{figure*}[!t]
	\centerline{\includegraphics[width=0.95\textwidth]{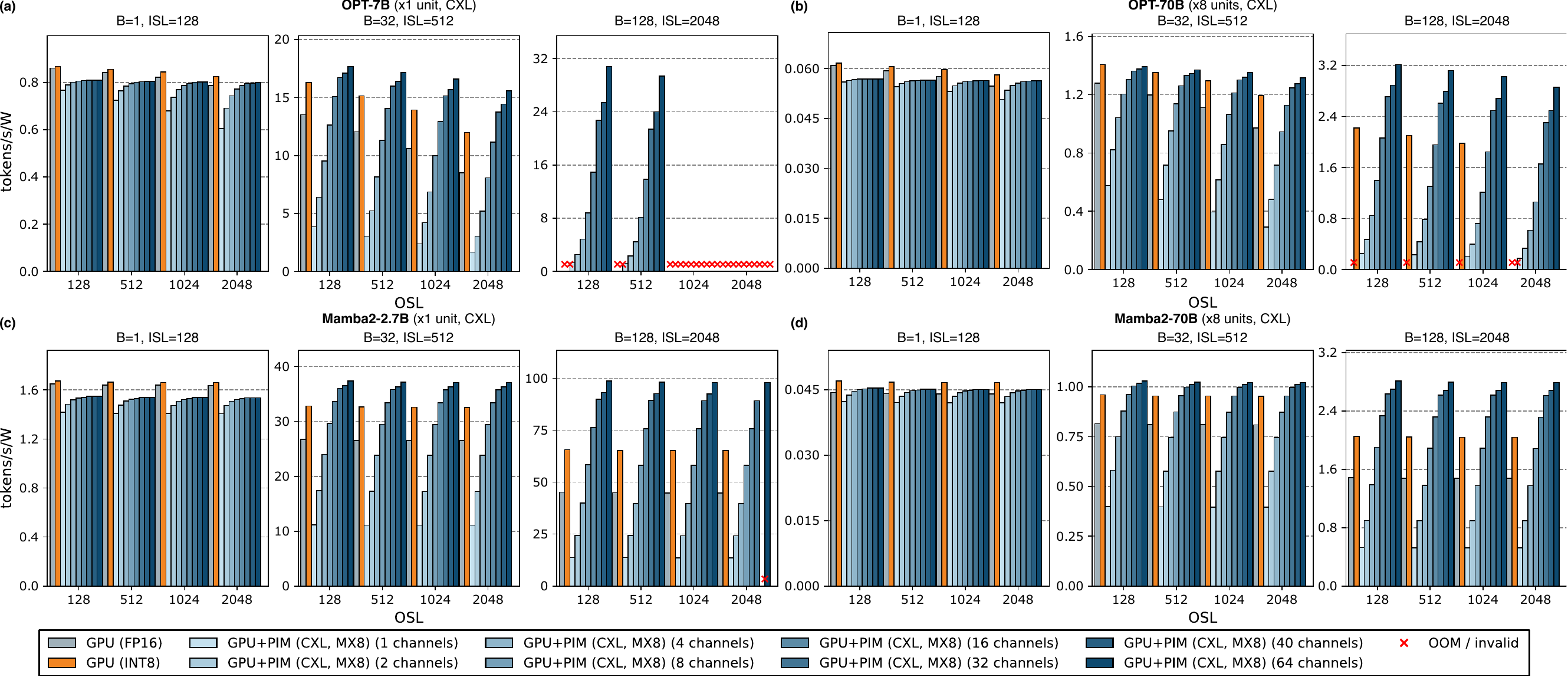}}
	\vspace{-0.5em}
	\caption{Decoding performance vs.\ channel count [1--64] with fixed pseudochannel/bg/banks (2,4,4) for (a) OPT-7B, (b) OPT-70B ($\times$8), (c) Mamba2-2.7B, (d) Mamba2-70B ($\times$8). GPU(s) vs.\ GPU+PIM (CXL). All configurations include static power. PIM capacity per unit scales with channel count, from $\sim$2\,GiB at 1 channel to $\sim$128\,GiB at 64 channels (OPT-70B/Mamba2-70B use 8 units, so total system capacity is 8$\times$ these values).}
	\label{fig:channels}
	\vspace{-1em}
\end{figure*}

We evaluate tokens/s/W across batch size (B), \acp{ISL}, and \acp{OSL}, with and without static power (Fig.~\ref{fig:static}). Including static power notably reduces relative tokens/s/W, especially at large B and ISL.
The 3.85$\times$ headline is \emph{relative overestimation} (largest where dynamic-only models predict the largest PIM gains, hence at Mamba2-2.7B, B=1), not absolute PIM benefit. At B=1, default-hierarchy PIM matches the GPU baseline; throughput-amortized gains emerge only at B$\geq$32 (a single-channel hierarchy recovers $\leq$4\% even at B=1; see \S\ref{sec:hierarchy}).
Trends are broadly consistent across model scales, though adding more units diminishes per-unit gains as added static power offsets throughput.
Integrated and co-located (CXL, NVLink) configurations achieve similar performance, suggesting that interconnect choice is secondary to static power management.
Despite the 2.5$\times$ NVLink3-vs-CXL bandwidth gap, transfer volume is small (PIM-offloaded kernels exchange only compact partial sums and KV slices); the \texttt{comm} component remains a small fraction of decode latency and energy across all three configurations, with static power dominating the envelope.
For the modeled decode offload strategy, interconnect bandwidth is not the dominant efficiency limiter.

Fig.~\ref{fig:breakdown} reports the power/latency breakdown for OPT-70B and Mamba2-70B under GPU+PIM (CXL). \emph{Static components (leakage, refresh, GPU idle) account for 66--69\% of total system power}, dominating the efficiency envelope. The high idle-GPU share arises from context-active board power that persists when no GPU kernels are executing, compounded by frequent synchronization/data-transfer phases of low utilization. Time spent on \texttt{gemv} is substantially larger for transformers than for SSM state updates, reflecting different memory access patterns: attention reads scale with sequence length, while SSM updates operate over fixed-size buffers.
Sensitivity analysis (Table~\ref{tab:sensitivity}) confirms robustness of the reported DRAM-PIM advantage: independently perturbing DRAM frequency, CXL bandwidth, GPU--GPU all-reduce latency, GPU context-active power, PIM-HBM background power, and PIM refresh energy, each across its full documented sensitivity range, places PIM's (CXL) tokens/s/W advantage over GPU (OPT-70B, B=28 -- the capacity-valid substitute for the submitted B=128 stress point at this ISL/OSL, which OOMs for GPU-FP16 -- ISL=OSL=2048, $\times$8 units) between 1.35$\times$ and 2.09$\times$ when static power is included (nominal 1.63$\times$). Excluding static power, the advantage (2.01$\times$, 4.198\,tok/s/W nominal) is \emph{exactly} invariant across every range tested.

\begin{table}[!t]
	\caption{Sensitivity analysis: PIM (CXL) tokens/s/W vs.\ GPU, OPT-70B, B=28, ISL=OSL=2048, $\times$8 units, incl.\ static power.}
	\label{tab:sensitivity}
	\centering
	\scriptsize
	\setlength{\tabcolsep}{3pt}
	\begin{tabular}{lcc}
		\toprule \toprule
		\textbf{Parameter}          & \textbf{Range}           & \textbf{tok/s/W (vs.\ GPU)}       \\
		\midrule
		GPU context-active power    & 54--105\,W               & 0.863--1.333 (1.35--2.09$\times$) \\
		PIM-HBM background power    & 2.10--8.41\,W            & 1.044--1.102 (1.63--1.72$\times$) \\
		DRAM frequency              & 0.8--1.2$\times$ nominal & 1.044--1.060 (1.63--1.66$\times$) \\
		CXL bandwidth               & 32--120\,GB/s            & 1.031--1.044 (1.61--1.63$\times$) \\
		PIM refresh energy/event    & 0.72--4.89\,nJ           & 1.044--1.056 (1.63--1.65$\times$) \\
		GPU--GPU all-reduce latency & 1.0--6.06\,$\mu$s        & 1.041--1.046 (1.63--1.64$\times$) \\
		\bottomrule \bottomrule
	\end{tabular}
\end{table}

\subsection{DRAM Hierarchy Co-optimization}\label{sec:hierarchy}
Given a fixed pseudochannel, bank groups, and banks (2, 4, 4), we vary the number of channels (Fig.~\ref{fig:channels}). Per-unit PIM capacity scales with channel count, from approximately 2\,GiB at 1 channel to 128\,GiB at 64 channels, matching the canonical 80\,GiB only at 40 channels; Fig.~\ref{fig:channels} therefore represents whole-unit provisioning, where capacity and parallel resources scale together with channel count, rather than an isolated channel-count experiment. Decoding performance is monotonically non-decreasing with channel count for every model and workload we evaluate, generally plateauing at high channel counts for low-batch workloads: more channels provide both more independent memory streams for row-locality-sensitive PIM kernels (\texttt{gemv}, \texttt{state\_update}).
At B=1, gains from added channels are modest (5--32\% depending on model and OSL); performance generally begins to plateau between 32 and 64 channels, depending on the model and sequence length. At B$\geq$32, every capacity-valid configuration for every model peaks at the largest channel count we evaluate (64), with gains of 140--2300\% over a single channel.

\begin{figure*}[!t]
	\centerline{\includegraphics[width=1\textwidth]{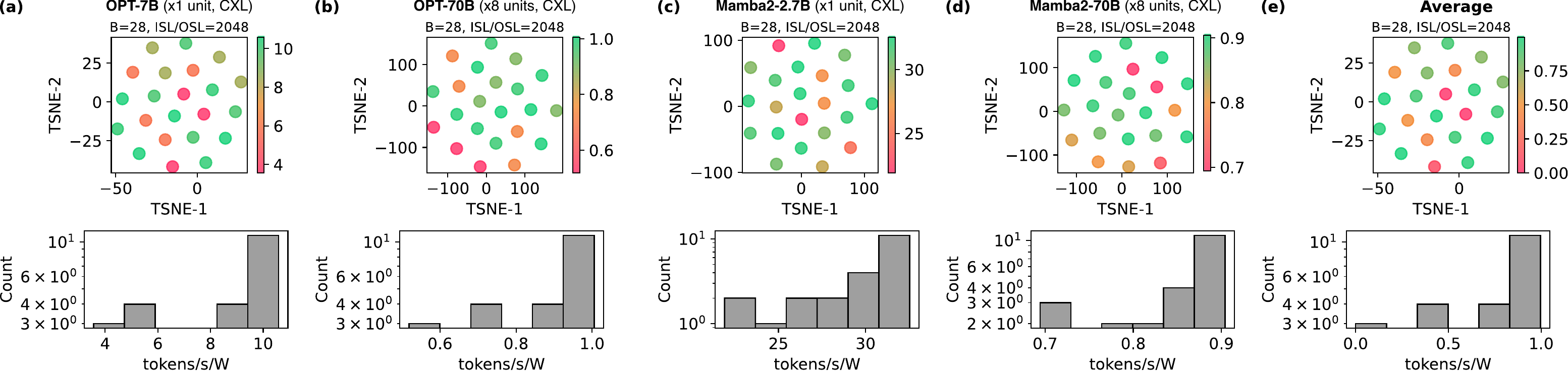}}
	\vspace{-0.5em}
	\caption{t-SNE (channels, pseudochannels, ACT4-group count) for the 22 (of 25) configurations in the extended fixed-capacity logical design space; 3 excluded as infeasible (Sec.~\ref{sec:hierarchy}). Physical bank count (16--5,120) and NBPU count (8--2,560) vary across the sweep; bank groups/banks fixed at (4,4); PIM capacity fixed at 80\,GiB per unit. (a-d) Raw tokens/s/W per model, each embedded independently; (e) normalized tokens/s/W. B=28, ISL/OSL=2048, GPU+PIM (CXL).}
	\label{fig:tsne}
	\vspace{-1em}
\end{figure*}

Further, in Fig.~\ref{fig:tsne}, we fix PIM capacity at 80\,GiB per unit and vary its distribution across channels/pseudochannels, holding bank groups/banks fixed at (4,4); the sweep is parametrized as (channels, pseudochannels, num\_act4\_groups) with their product fixed at 320, where num\_act4\_groups is the logical count of 4-bank ACT4 activation groups per pseudochannel (distinct from the fixed physical bank-group count of 4) that this sweep varies to hold per-unit capacity fixed while the physical bank count changes. Of 25 mathematically distinct triples, 3 require a refresh-barrier latency exceeding the simulator's fixed refresh interval for any workload and cannot complete.
All four models agree closely on the best configuration -- (40,2,4) and (80,1,4) both reach within 0.3\% of each model's own peak tokens/s/W; however, they differ sharply in sensitivity to a poor choice. The worst configuration, (2,2,80), retains only 33.6\%/51.6\% of peak for OPT-7B/OPT-70B versus 67.3\%/76.8\% for Mamba2-2.7B/Mamba2-70B.

\begin{figure}[!t]
	\centerline{\includegraphics[width=0.99\columnwidth]{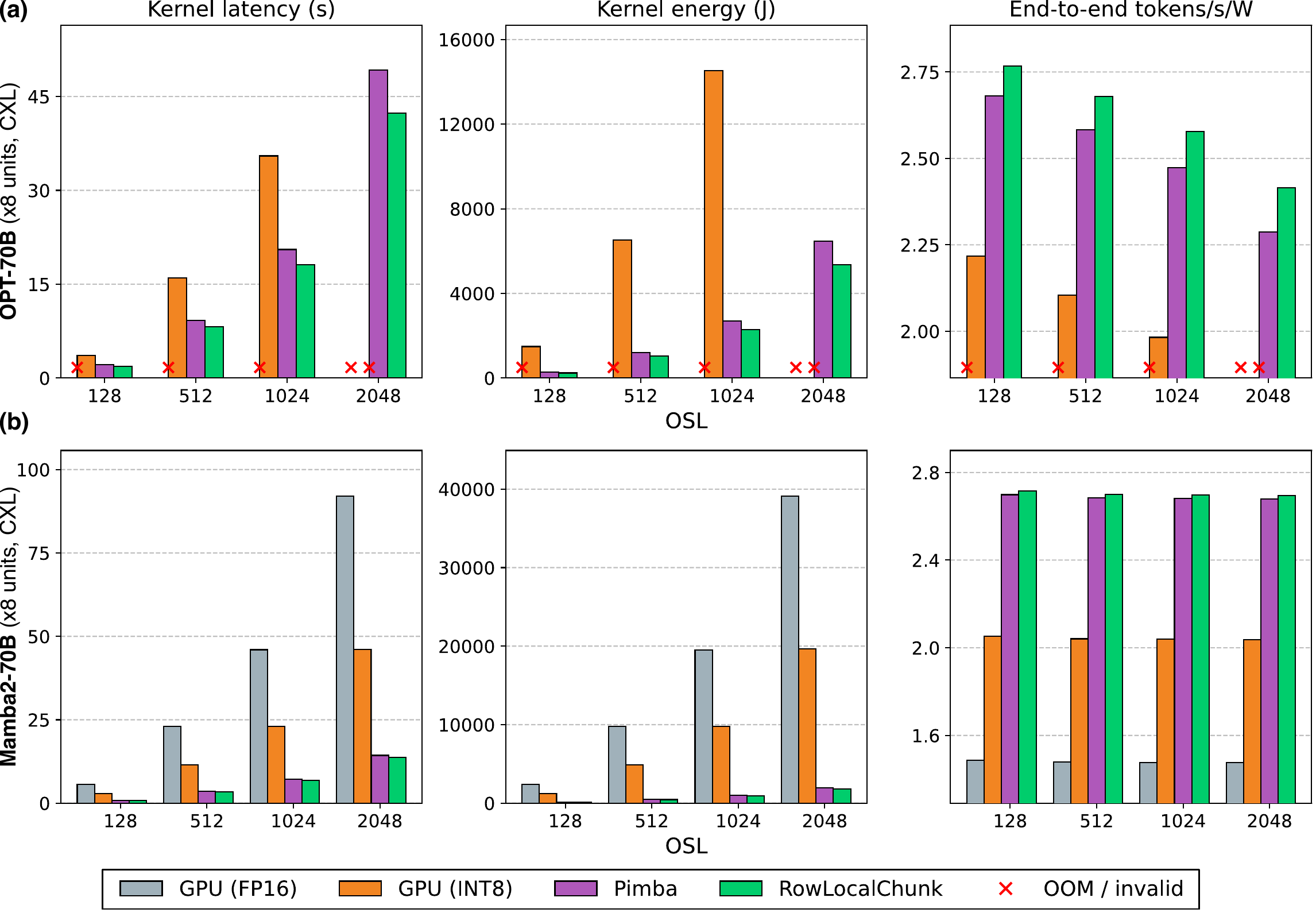}}
	\vspace{-0.5em}
	\caption{Comparison of Pimba and RowLocalChunk for OPT-70B (top) and Mamba2-70B (bottom). Columns show PIM-kernel latency, PIM-kernel energy, and end-to-end tokens/s/W. Crosses indicate OOM/invalid configurations.}
	\label{fig:mapping}
\end{figure}

\subsection{Workload Mapping Evaluation}
For both Pimba and RowLocalChunk (\S\ref{sec:mapping}, Algorithm~\ref{alg:genmap_compact}), DRAM commands are interleaved across banks/pseudochannels/channels, so locality improvements are not masked by controller-level parallelism (Fig.~\ref{fig:mapping}).
RowLocalChunk reduces kernel latency by 4.4--14.0\% and kernel energy by 6.3--17.4\% across evaluated OSLs by cutting activate/precharge cycles for adjacent chunks.
End-to-end decode gain is modest (0.6--5.6\%) because mapping governs only the PIM portion: PIM kernels are a fraction of total decode energy once GPU compute, communication, and static idle power are included (Fig.~\ref{fig:breakdown}), compressing kernel-level gains.

\section{Conclusion}
Through systematic evaluation of heterogeneous DRAM-PIM-GPU systems across LLM decode workloads, we
established three design principles. 
First, static power can dominate the efficiency calculus and must be included in system-level evaluations. Second, decoding performance is monotonically non-decreasing with channel count across all evaluated models and workloads, generally plateauing at high channel counts for low-batch workloads; under a fixed-capacity sweep, all models instead share a common near-optimal region. Third, workload mapping provides bounded gains and is not the dominant bottleneck; significant efficiency gains require system-wide co-optimization.
This motivates architectural reconfigurability.

\vspace{-0.5em}
\bibliographystyle{IEEEtran}
\bibliography{references}

@InProceedings{DarvishRouhani2023,
  author     = {Darvish Rouhani and others},
  booktitle  = {Proceedings of the 50th Annual International Symposium on Computer Architecture},
  title      = {{With Shared Microexponents, A Little Shifting Goes a Long Way}},
  year       = {2023},
  month      = June,
  publisher  = {ACM},
  collection = {ISCA ’23},
  doi        = {10.1145/3579371.3589351},
}

@InProceedings{Kim2025,
  author     = {Kim, Wonung and others},
  booktitle  = {Proceedings of the 58th IEEE/ACM International Symposium on Microarchitecture},
  title      = {{Pimba: A Processing-in-Memory Acceleration for Post-Transformer Large Language Model Serving}},
  year       = {2025},
  month      = Oct,
  pages      = {292--307},
  publisher  = {ACM},
  collection = {MICRO 2025},
  doi        = {10.1145/3725843.3756121},
}

@InProceedings{He2025,
  author     = {He, Yintao and others},
  booktitle  = {Proceedings of the 30th ACM International Conference on Architectural Support for Programming Languages and Operating Systems, Volume 2},
  title      = {{PAPI: Exploiting Dynamic Parallelism in Large Language Model Decoding with a Processing-In-Memory-Enabled Computing System}},
  year       = {2025},
  month      = Mar,
  pages      = {766--782},
  publisher  = {ACM},
  collection = {ASPLOS ’25},
  doi        = {10.1145/3676641.3716009},
}

@InProceedings{Gu2025,
  author     = {Gu, Yufeng and others},
  booktitle  = {Proceedings of the 30th ACM International Conference on Architectural Support for Programming Languages and Operating Systems, Volume 2},
  title      = {{PIM Is All You Need: A CXL-Enabled GPU-Free System for Large Language Model Inference}},
  year       = {2025},
  month      = Mar,
  pages      = {862--881},
  publisher  = {ACM},
  collection = {ASPLOS ’25},
  doi        = {10.1145/3676641.3716267},
}

@InProceedings{Park2024,
  author     = {Park, Jaehyun and others},
  booktitle  = {Proceedings of the 29th ACM International Conference on Architectural Support for Programming Languages and Operating Systems, Volume 2},
  title      = {{AttAcc! Unleashing the Power of PIM for Batched Transformer-based Generative Model Inference}},
  year       = {2024},
  month      = Apr,
  pages      = {103--119},
  publisher  = {ACM},
  collection = {ASPLOS ’24},
  doi        = {10.1145/3620665.3640422},
}

@Article{Luo2023,
  author    = {Luo, Haocong and others},
  journal   = {IEEE Computer Architecture Letters},
  title     = {{Ramulator 2.0: A Modern, Modular, and Extensible DRAM Simulator}},
  year      = {2024},
  issn      = {2473-2575},
  month     = Jan,
  number    = {1},
  pages     = {112--116},
  volume    = {23},
  doi       = {10.1109/lca.2023.3333759},
  publisher = {IEEE},
}

@Article{Choquette2021,
  author    = {Choquette, Jack and Gandhi, Wishwesh and Giroux, Olivier and Stam, Nick and Krashinsky, Ronny},
  journal   = {IEEE Micro},
  title     = {{NVIDIA A100 Tensor Core GPU: Performance and Innovation}},
  year      = {2021},
  issn      = {1937-4143},
  month     = Mar,
  number    = {2},
  pages     = {29--35},
  volume    = {41},
  doi       = {10.1109/mm.2021.3061394},
  publisher = {IEEE},
}

@inproceedings{Steiner2025,
author = {Steiner, Lukas and others},
title = {{DRAMPower 5: An Open-Source Power Simulator for Current Generation DRAM Standards}},
year = {2025},
isbn = {9798400714719},
publisher = {ACM},
doi = {10.1145/3721848.3721850},
booktitle = {Proceedings of the Rapid Simulation and Performance Evaluation for Design Workshop},
pages = {8–16},
numpages = {9},
location = {
},
series = {RAPIDO '25}
}

@misc{Zhang2022,
      title={{OPT: Open Pre-trained Transformer Language Models}}, 
      author={Susan Zhang and others},
      year={2022},
      eprint={2205.01068},
      archivePrefix={arXiv},
      primaryClass={cs.CL},
      url={10.48550/arXiv.2205.01068}, 
}

@inproceedings{Dao2024,
author = {Dao, Tri and Gu, Albert},
title = {{Transformers are SSMs: generalized models and efficient algorithms through structured state space duality}},
year = {2024},
publisher = {JMLR.org},
booktitle = {Proceedings of the 41st International Conference on Machine Learning},
articleno = {399},
numpages = {31},
location = {Vienna, Austria},
series = {ICML'24}
}

@Article{Wang2024,
  author    = {Wang, Junpeng and others},
  journal   = {IEEE Transactions on Computer-Aided Design of Integrated Circuits and Systems},
  title     = {{NicePIM: Design Space Exploration for Processing-In-Memory DNN Accelerators With 3-D Stacked-DRAM}},
  year      = {2024},
  issn      = {1937-4151},
  month     = May,
  number    = {5},
  pages     = {1456--1469},
  volume    = {43},
  doi       = {10.1109/tcad.2023.3342605},
  publisher = {Institute of Electrical and Electronics Engineers (IEEE)},
}

@Misc{khan2024landscapecomputenearmemorycomputeinmemoryresearch,
  author    = {Khan, Asif Ali and others},
  title     = {{The Landscape of Compute-near-memory and Compute-in-memory: A Research and Commercial Overview}},
  year      = {2024},
  copyright = {Creative Commons Attribution 4.0 International},
  url       = {10.48550/arxiv.2401.14428},
  publisher = {arXiv},
}

@Article{Ghose2018,
  author    = {Ghose, Saugata and others},
  journal   = {Proceedings of the ACM on Measurement and Analysis of Computing Systems},
  title     = {{What Your DRAM Power Models Are Not Telling You: Lessons from a Detailed Experimental Study}},
  year      = {2018},
  issn      = {2476-1249},
  month     = Dec,
  number    = {3},
  volume    = {2},
  doi       = {10.1145/3224419},
  publisher = {Association for Computing Machinery (ACM)},
}
\end{document}